\documentclass[pdflatex,sn-mathphys]{sn-jnl}% Math and Physical Sciences Reference Style
\usepackage{siunitx}
\usepackage{enumerate}
\usepackage{amsmath,latexsym,amssymb}
\usepackage[export]{adjustbox}
\usepackage{algpseudocode}
\usepackage{listings}
\usepackage{ulem}

\definecolor{code-green}{rgb}{0,0.6,0}
\definecolor{code-gray}{rgb}{0.5,0.5,0.5}
\definecolor{code-mauve}{rgb}{0.58,0,0.82}
\jyear{2021}%

\theoremstyle{thmstyleone}%
\theoremstyle{thmstyletwo}%

\theoremstyle{thmstylethree}%

\newcommand{\ssh}{S_{\rm sh}}
\newcommand{\sho}{S_{\rm ho}}
\newcommand{\snr}{{\rm SNR}}

\begin{document}
	
	\title[]{
		Operating Fiber Networks in the Quantum Limit
		%Data Transmission in the Quantum Limit
		%Data Networks in the Quantum Limit
	}
	
	
	\author*[1]{\fnm{Janis} \sur{N\"otzel}}\email{janis.notzel@tum.de}
	\equalcont{These authors contributed equally to this work.}
	
	\author*[2]{\fnm{Matteo} \sur{Rosati}}\email{matteo.rosati@uab.cat}
	\equalcont{These authors contributed equally to this work.}
	
	\affil[1]{\orgdiv{Emmy-Noether Gruppe Theoretisches Quantensystemdesign\\Lehrstuhl f\"ur Theoretische Informationstechnik}\\ \orgname{
			Technische Universit\"at M\"unchen}} %\orgaddress{\street{Street}, \city{City}, \postcode{100190}, \state{State}, \country{Country}}}

\affil[2]{\orgdiv{Departament de F\'{\i}sica: Grup d'Informaci\'{o} Qu\`{a}ntica},\\ \orgname{Universitat Aut\`{o}noma de Barcelona}, \orgaddress{\city{ Bellaterra (Barcelona)}, \postcode{ES-08193}, \country{Spain}}}

\abstract{%abstract
	We consider all-optical network evolution from a quantum perspective. We show that a use of optimal quantum receivers allows an estimated $55\%$ decrease in energy consumption of all-optical amplifiers in network configurations that are typical today. We then compare data transmission capacities of quantum receivers with today's technology operating within the boundaries set by Shannon. We find that quantum receiver technology allows for a logarithmic scaling of the system capacity with the baud-rate, while Shannon-type systems are limited by the transmit power. Thus a natural quantum limit of classical data transmission emerges. Based on the above findings we argue for a new approach to optical communication network design, wherein in-line amplifiers are replaced by novel fiber supporting high spectral bandwidth to allow for noiseless data transmission in the quantum limit.

}

\keywords{Optical networks, Green communication, Quantum communication, Joint detection receiver}

\maketitle

\section{Introduction}\label{sec1}
With this work we point out one possible near-term contribution of quantum information processing techniques to existing and near-term data transmission networks, focusing in particular on fiber networks. For the existing networks we focus on quantum receiver technology as a tool for reducing energy consumption in amplifiers. For future networks we explain how amplifiers could be avoided by utilizing the scaling law of the fiber link capacity with the baud-rate when quantum detection methods are used.
%The analysis carried out in this work is based on the communication models introduced by Shannon \cite{Shannon1948b} and Holevo \cite{Holevo1973}, and should be understood as a motivation for more in-depth analysis of quantum data transmission techniques beyond the scope of such basic models.
% now old text
\subsection{Energy Savings in Current Networks}
The progressive development of fiber networks is a driver of productivity growth in modern societies \cite{bay16}. As a result, network electricity consumption has been estimated as $\approx1.7\%$ of the global consumption, with a growth rate of $\approx10\%$ per year~\cite{VanHeddeghem2014} in 2012. Thus, energy efficiency has become a key network design principle and one of the main motivations to switch from electrical to optical information-processing in networks, e.g., via the introduction of integrated-photonics circuits~\cite{kil16}.
Interestingly, such circuits are also one of the major candidate platforms for quantum information processing (QIP), a field which has long promised to boost the capabilities of classical information technology~\cite{Elshaari2020,Wang2020}.

However, the potential advantages of QIP for classical data transmission reported so far can be described as ``large gains at small practical value", e.g., the reduction of peak signal energy for deep-space communication and the increase of transmission rate in short-distance communication~\cite{Waseda10,Waseda11,Guha11,Rosati16c,Antonelli2014,Jarzyna2019a}. 

In this work we change such perspective, proving that QIP can provide a significant reduction of energy consumption in  classical-data-transmission systems that employ optical amplification, until even the hypothetical removal of the amplifiers.
Optical amplification is a widespread system design, wherein amplifiers are placed along the path to counteract propagation losses that progressively reduce the signal's strength, ensuring that information can be reliably decoded by the receiving side~\cite{Tucker2011a,Tucker2011b,Pillai2014,Lundberg2017}. Such amplifiers consume a considerable amount of energy, e.g., for a \SI{1100}{\km}-link their cost is estimated as $\approx13\%$ of the total energy consumption~\cite{Pillai2014}; hence 
%our main focus is on reducing the amplification energy budget.
a reduction of their energy budget is a worthy objective.

In this setting, we first show that a current communication system employing a quantum optimal joint-detection receiver (OJDR)\cite{Schumacher97,Hausladen1996,Guha2012,Wilde2013a,Giovannetti2011,Giovannetti2012,Rosati16b} can transfer the same amount of bits of a classical optimal single-symbol receiver (OSSR) by spending less energy per signal, thanks to a net reduction of the amplification costs. %{\color{red} In a second step, we investigate the scaling of the system capacities with the baud-rate to highlight a potential future development trend which is based on use of the OJDR. }
The difference between these two receiver designs lies in the way they manipulate the signals: the OSSR performs individual operations on each received signal, thereby immediately producing a classical measurement outcome per pulse; in contrast, the OJDR can perform any collective operation on a long sequence of received signals prior to producing its measurement outcome.
Therefore, our results harness the OJDR's superior capability of reading off the transmitted data, which is particularly pronounced for signals with low added noise, by extracting the same amount of information from signals of lower energy with respect to an OSSR.

We start by providing a heuristic motivation for the reduction of the amplification costs, based on a striking result: in a quantum communication system employing OJDR, the strategy of turning on all the available amplifiers can be sub-optimal, contrarily to what happens for an OSSR. This property, which favours lower added noise at the cost of a smaller signal amplification, can be used to reduce the amplification energy budget while increasing the communication rate.

We then refine our analysis by considering tunable amplifier gains, which measure the strength of the amplification process and can be related to its energy cost. We formulate the problem of minimizing the total energy cost with respect to the gain profile with OJDR, while attaining a larger communication rate than the OSSR. We provide an algorithm to solve this problem and employ it to predict the maximum reduction of energy consumption in a wide range of distances and numbers of amplifiers. 

We show that energy savings arising from the use of an OJDR can be as large as $\approx100\%$. However, this happens only in settings where the communication system benefits very little from the use of amplifiers in the first place. Instead, in the practically relevant setting where the amplifiers enhance the data-rate at least by a factor of 2, we identify situations where the use of an OJDR allows record energy savings up to $\approx55\%$.%$57\%$.

To the best of our knowledge, the minimization of energy consumption of amplifiers through OJDR technology has been analyzed only in a couple of recent works, though from different perspectives: \cite{Jarzyna2019a} reported a constant increase of the communication rate using the OJDR instead of the OSSR on an optically-amplified multi-span link; %by up to $\approx \SI{1.44}{bits/(s\cdot Hz)}$
whereas \cite{Antonelli2014} studied the energy-efficiency of a system employing continuous amplification in the limit where the OSSR approximates the OJDR, which fails to capture the full advantage of the OJDR. %(see 
We combine both approaches to quantify the energy-efficiency advantage of the OJDR on a multi-span optically-amplified link.
\subsection{Future Quantum Data Networks}
Finally, we provide a theoretical justification for a novel approach to future data transmission network design 
and detail the first implementable steps to its realization.
%.
To this purpose, we point out the strikingly different growth of data transmission capacities of both OSSR and OJDR system-design approaches: under any power limit and any thermal noise level, measured in photons per second, the link capacity grows unbounded with a logarithmic dependence on the baud-rate when an OJDR is used as the receiver. In sharp contrast, the link capacity is upper-bounded in terms of the transmit power when an OSSR is used. Our technological hypotheses can be tested with novel system design concepts resting on the works of Guha et al. \cite{Guha11,Guha2012} and follow-up works of Rosati et al. \cite{Rosati16c} and Klimek et al. \cite{Klimek2015}. For the optical transmission medium, we suggest hollow-core fiber \cite{broadband-hollow-core,broadband-hollow-core-communications}.

\section{Results}\label{sec2}

\subsection{Communication System Model}\label{subsec:models-and-notation}
\begin{figure}[!t]
	\centering
	\includegraphics[trim={0 10cm 0 9cm},clip,width=\textwidth]{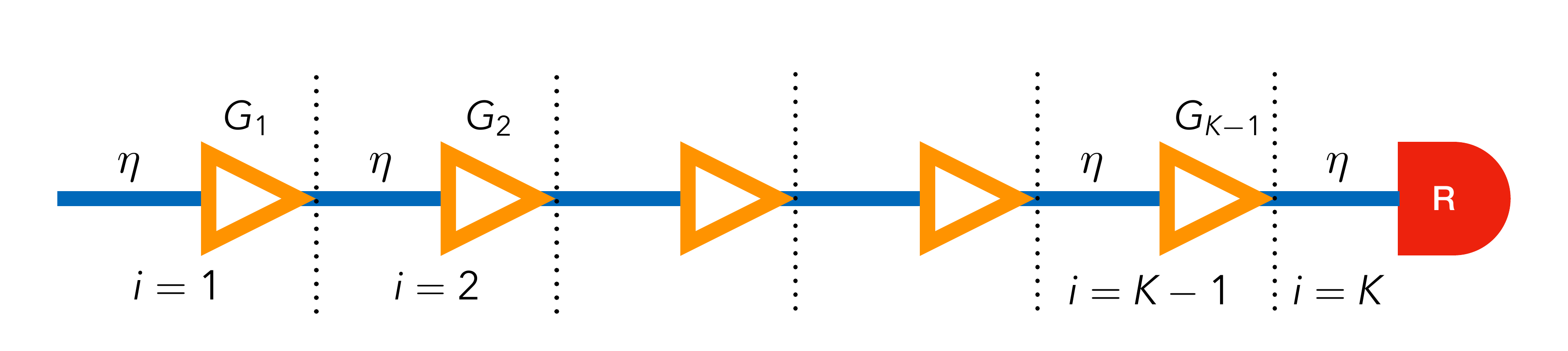}
	\caption{Depiction of the communication system studied: the optical-fiber link is divided into $K$ segments interleaved by optical amplifiers. Classical information is encoded in a quantum state and sent on the channel. Transmission on each segment $i$ is modelled by a pure-loss bosonic channel of attenuation $\eta$ followed by a quantum-limited bosonic amplifer channel of gain $G_i$. The last segment does not benefit from amplification and it is followed by a receiver $R$ that extracts the classical message from the received state. We show that, by using a quantum receiver, one can reduce the amplifier gains and thus spare significant amounts of energy to transfer the same amount of information with respect to a classical receiver.}\label{fig:comm-line}
\end{figure}

Consider a multi-span transmission line comprising an optical-fiber link of length $L$ and $K$ optical-amplifier modules placed at equally spaced intervals of length $L/K$\footnote{The techniques we develop can be applied to amplifiers with arbitrary spacing. Since this does not mark any qualitative change in our results, we restrict to equal spacing for simplicity.} (see Figure \ref{fig:comm-line}). Each of these components is modelled by a quantum bosonic Gaussian channel, taking as input a quantum state of the electromagnetic field. For such channels, the maximum information transmission rate can be attained by encoding classical information into sequences of optical coherent states~\cite{Giovannetti2014,Giovannetti2015,Mari2014} and performing a collective quantum measurement of the received sequences, to recover the classical message~\cite{Holevo1998c,Schumacher97,Hausladen1996,Guha2012,Wilde2013a,Giovannetti2011,Giovannetti2012,Rosati16b}.

For our purposes, the action of each transmission-line component can be described in terms of input-output relations for the mean energy of the field comprised of signal and noise \footnote{Throughout the main text, except in Results~\ref{subsec:numerics}, we set $\hbar\omega=1$ for some fixed signal frequency $\omega$, effectively equating the mean energy of the field with its mean number of photons.}:
\begin{enumerate}[(i)]
	\item transmission on an optical-fiber segment is modelled by a pure-loss channel,
	\begin{equation}
		n \mapsto \eta\cdot n,
	\end{equation}
	with $\eta:=e^{-\alpha \frac L K}$ and $\alpha\geq0$ a system-specific coefficient;
	\item optical amplification is modelled by a quantum-limited amplifier channel,
	\begin{equation}
		n \mapsto G\cdot n + G-1,
	\end{equation}
	with $G\geq 1$ the amplifier gain and $G-1$ the minimum photon-noise addition allowed by quantum mechanics.
\end{enumerate}
The action of the entire transmission line is then obtained by repeatedly applying these two channels: at the end of the $i$-th segment, the input  plus noise  to the next segment has energy $\tau_i\cdot n + \nu_i$, with
\begin{equation}\label{eqn:definition-of-tau-and-nu}
	\tau_i= G_i \cdot \eta \cdot \tau_{i-1}, \quad \nu_i=G_i\cdot \eta \cdot \nu_{i-1} + G_i -1,
\end{equation}
and initial values $\tau_0=1$, $\nu_0=0$. The resulting coefficients for the entire transmission line are $\tau:=\tau_K$, $\nu:=\nu_K$.

Communication performance is quantified via the spectral efficiency (SE), i.e., the maximum number of bits transmittable per unit time and frequency, which in our narrowband case coincides with the channel capacity. Importantly, depending on the receiver's capabilities, this quantity can take two different values\footnote{Throughout the article the logarithms are taken in base $2$.}: the Shannon SE~\cite{Shannon1948b}, attainable by performing heterodyne detection on each received signal,
\begin{equation}
	\ssh\left(n,\{G_i\}_{i=1}^K\right):=\log\left(1+\frac{\tau\cdot n}{1+\nu}\right),\label{eqn:shannon-S}
\end{equation}
and the Holevo SE~\cite{Holevo1973,Holevo1998c}, attainable by performing a collective quantum measurement (i.e., the OJDR for this channel) on the received sequences,
\begin{equation}
	\sho\left(n,\{G_i\}_{i=1}^K\right):=g(\tau\cdot n + \nu) - g(\nu),\label{eqn:holevo-S}
\end{equation}
where $g(x):=(x+1)\log(x+1) - x\log x$.

We stress that the realization of a collective measurement that attains \eqref{eqn:holevo-S} is still an open problem in quantum information theory, with several theoretical proposals that we refer to as OJDR~\cite{Schumacher97,Hausladen1996,Guha2012,Wilde2013a,Giovannetti2011,Giovannetti2012,Rosati16b}. Specifically, the all-optical Hadamard receiver can approximate the OJDR for low received signal energy~\cite{Guha11,Rosati16c} and no noise, but it is unknown whether a broader class of optical receivers can implement the OJDR in general~\cite{Rosati2017,Rosati17c,Nair2014,Takeoka14,Bilkis2020}.
On the other hand, \eqref{eqn:shannon-S} is attained with an explicit receiver design, which coincides with the OSSR for this channel in the practically relevant regime $\tau\cdot n\gtrsim 2$~\cite{Takeoka14,Holevo2019}. For completeness, in Methods~\ref{appendix:homodyne} we study the performance of the homodyne receiver as well, which coincides with the OSSR for $\tau\cdot n\lesssim 2$, observing an energy-advantage of OJDR also in this case. 

\subsection{Maximizing spectral efficiency: more data with less amplification}\label{sec:op-se}
We are now interested in maximizing the SE's (\ref{eqn:shannon-S},\ref{eqn:holevo-S}) with respect to the amplifier gains. Following the approach taken in \cite{Jarzyna2019a}, we assume that an energy constraint is imposed on the entire communication link, such that the energy of the field at any point along the link stays below a threshold:
\begin{align}\label{eqn:power-constraint}
	\tau_i\cdot n + \nu_i \leq n_{\max}\ \forall i=1,\ldots,K.
\end{align}
The value of $n_{\max}$ can be determined by practical constraints; in particular we will focus on the case where the sender produces signals of maximum energy, i.e., $n=n_{\max}$. In the following we study the problem of spectral-efficiency-optimal gain selection (SEGS):
\begin{equation}\begin{aligned}\label{eqn:mathematical-problem-statement}
		%\mathbf{Problem:\ }&\mathrm{Energy-optimal\ gain\ selection\ (EGS)}\\
		S^{\rm op}(n):=\underset{\{G_i\}_{i=1}^{K}\in\mathbb R^{K}}{\mathrm{maximize}}\ & S(n,\{G_i\}_{i=1}^{K})\\
		\mathrm{subject\ to:}\ &G_K=1, 1\leq G_i\leq G_{\max}\ \forall i,\\  &\tau_i\cdot n + \nu_i \leq n_{\max}\ \forall i=1,\ldots,K,
\end{aligned}\end{equation}
where $S$ is either the Shannon or Holevo SE and the corresponding problems are called Shannon-SEGS and Holevo-SEGS. 
Let us note that the last amplifier can always be turned off, i.e., $G_K=1$. Indeed, this amplifier simply constitutes an additional channel acting after the signal is received; hence, by the data-processing inequality, it cannot increase the overall spectral efficiency. 

For the Shannon-SEGS, the optimal amplification strategy is simply to turn on all the remaining amplifiers at the maximum value of the gain allowed by the energy constraint \eqref{eqn:power-constraint}:
\begin{equation}
	G_{i<K}=G_{\max} :=\frac{1+n}{1+\eta\cdot n}\quad \forall i<K.
\end{equation}
This follows from the fact that the Shannon SE \eqref{eqn:shannon-S} is a monotonically increasing function of $G_i$ for all $i<K$ (see Methods~\ref{appendix:shannon-segs}).
The resulting optimal Shannon SE is
\begin{equation}\label{eqn:shannon-op}
	\ssh^{\rm op}(n)=\log\left(1+\frac{\tau_{\max}\cdot n}{1+(\eta-\tau_{\max})\cdot n}\right), \quad \tau_{\max}=\eta^K\cdot (G_{\rm max})^{K-1}.
\end{equation}

On the other hand, the Holevo-SEGS amplification strategy can be considerably different, due to the presence of a quantum OJDR. In this case, the optimization depends non-trivially on the system parameters $\eta$, $K$ and on the input signal energy $n$. Still, it can be shown that the optimal sequence of gains is non-increasing with $i$; indeed, if two gains are in increasing order, switching them always decreases the noise $\nu$ in the overall channel without changing its amplification coefficient $\tau$ (see Methods~\ref{appendix:decreasing-gains-optimal}). This property, together with the well-known fact that the Holevo SE is never smaller than the Shannon SE~\cite{Holevo1973,Holevo1998c} leads us to the conclusion that 
\begin{equation}
	\sho^{\rm op}(n)\geq\sho(n,\{G_{i<K}=G_{\max},G_K=1\})\geq\ssh^{\rm op}(n).
\end{equation}
Moreover, in Methods~\ref{appendix:holevo-segs} we prove that the first inequality can be strict, i.e., there exists a whole range of system parameter values such that setting maximum gains results in a sub-optimal value of $\sho$ and, in fact, a certain number of amplifiers can even be turned off. 

We thus conclude that, in general, starting from a fully-amplified link with OSSR, the introduction of an OJDR allows to decrease amplifier gains while increasing the spectral efficiency. In turn, since smaller gains imply a smaller energy consumption to operate the amplifiers, the OJDR allows to spare a certain amount of energy. This motivates taking a comparative look at energy-optimal gain selection for OJDR versus OSSR.

\subsection{Minimizing amplifier energy consumption:\\Problem definition}
Every amplifier needs to use up at least the amount of energy that it adds to the incoming field, hence the total amplification energy consumption is at least
\begin{align}\label{def:energy-consumption}
	E(n,\{G_i\}_{i=1}^K) = \sum_{i=1}^{K}(G_i-1)(\eta(\tau_{i-1} n+\nu_{i-1})+1).
\end{align}
A fully-amplified link with OSSR attains SE $\ssh^{\rm op}(n)$ with an energy consumption of $E_{\rm sh}:=(K-1)(1-\eta)n$. Using the former as a baseline, we can then define the following energy-optimal gain selection (EGS) problem for a transmission link with OJDR:
\begin{equation}\begin{aligned}\label{eqn:mathematical-problem-statement1}
		%\mathbf{Problem:\ }&\mathrm{Energy-optimal\ gain\ selection\ (EGS)}\\
		E_{\rm egs}:=\underset{\{G_i\}_{i=1}^{K}\in\mathbb R^{K}}{\mathrm{minimize}}\ & E(n,\{G_i\}_{i=1}^{K})\\
		\mathrm{subject\ to:}\ &G_K=1, 1\leq G_i\leq G_{\max}\ \forall i,\\  &\sho(n,\{G_i\}_{i=1}^{K}) \geq \ssh^{\rm op}(n).
\end{aligned}\end{equation}
Importantly, note that the energy needed to provide a given gain $G_i$ depends on all previous gains $G_{j<i}$ and on the attenuation per segment $\eta$. In particular, as the noise added by previous amplifiers increases, so does the energy cost for providing a certain gain. This implies that monotonically decreasing gains are not necessarily optimal for EGS.

We therefore also consider a relaxation of EGS (REGS): 
\begin{equation}\begin{aligned}\label{eqn:mathematical-problem-statement2}
		%\mathbf{Problem:\ }&\mathrm{Relaxed\ energy-optimal\ gain\ selection\ (REGS)}\\
		E_{\rm regs}:=\underset{\{G_i\}_{i=1}^{K}\in\mathbb R^{K}}{\mathrm{minimize}}\ & E'(n,\{G_i\}_{i=1}^{K})\\
		\mathrm{subject\ to:}\ &G_K=1, 1\leq G_i\leq G_{\max}\ \forall i,\\  &\sho(n,\{G_i\}_{i=1}^{K}) \geq \ssh^{\rm op}(n),
\end{aligned}\end{equation}
Here the energy cost for each amplifier is overestimated as the maximum one, via \eqref{eqn:power-constraint}, i.e., \begin{equation}E'(n,\{G_i\}_{i=1}^{K})= \sum_{i=1}^{K}(G_i-1)(\eta\cdot n+1),\end{equation}
and it does not depend on the previous gains. Therefore, in this case, it is clear that non-increasing gains can always attain the optimum of REGS and, at the same time, maximize the SE, as discussed in Sec.~\ref{sec:op-se}.

While non-increasing gains already provide reductions in energy consumption when using an OJDR, the subtle differences between problems EGS and REGS show that optimum gain selection for quantum communication networks is an important nontrivial aspect of quantum network design by itself.

\subsection{Minimizing amplifier energy consumption: Numerical Results}\label{subsec:numerics}
\begin{figure}[!t]
	\centering
	\begin{minipage}[t]{.828\textwidth}
		\includegraphics[width=\textwidth,valign=t]{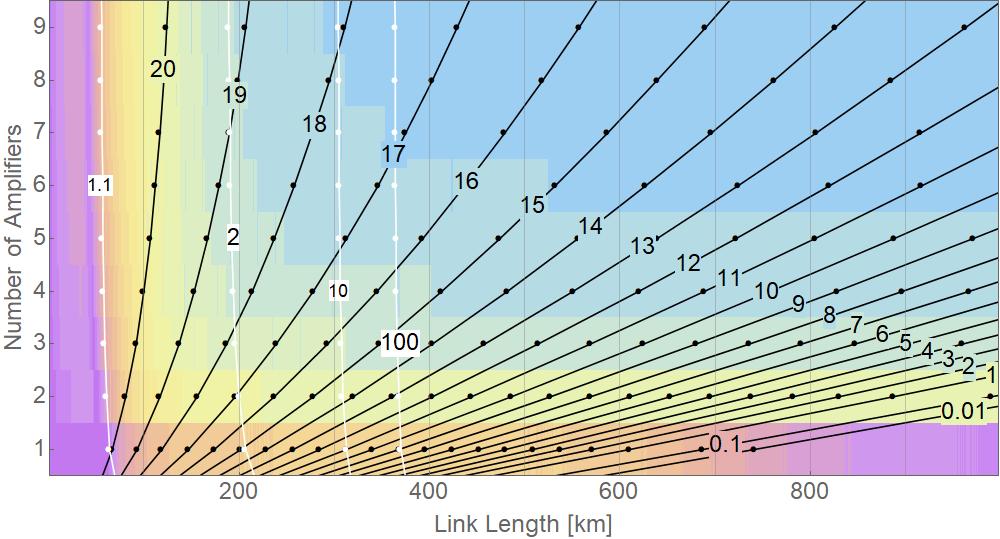}
		\includegraphics[width=\textwidth,valign=t]{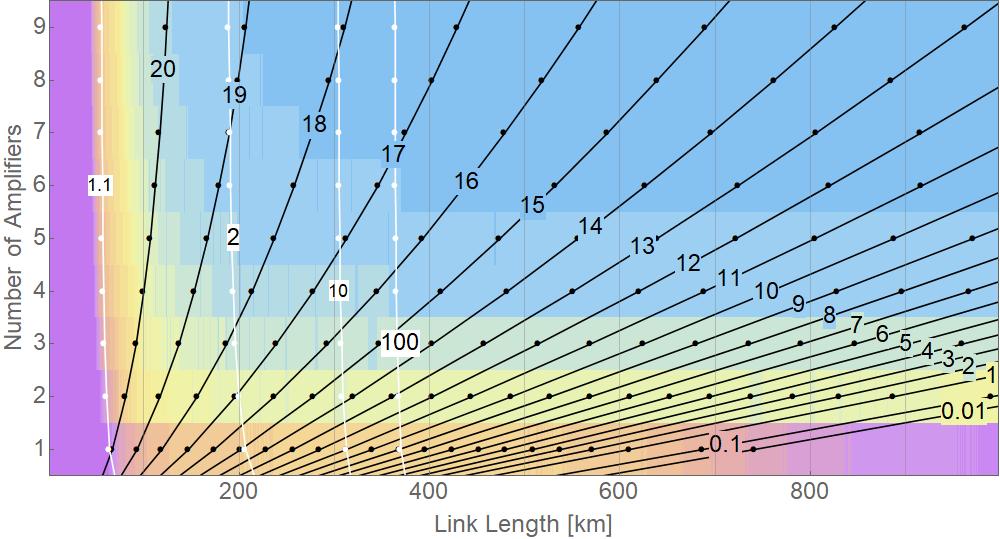}
	\end{minipage}
	\begin{minipage}[t]{.164\textwidth}
		\includegraphics[width=\textwidth,valign=t]{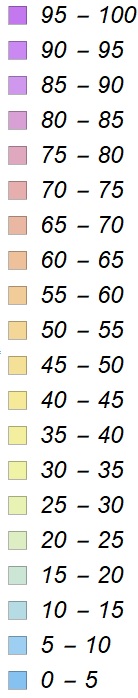}
	\end{minipage}
	\caption{Displayed is the percentage of energy used by the in-line amplifiers of links operating an OJDR relative to the same link with OSSR, as a function of link length and number of amplifiers. The top (bottom) image shows energy savings calculated according to problem EGS (REGS).
		The photon number is set to $n=10^7$, corresponding to a maximum signal power of $\mathrm{100mW}$. SE is constant along black lines. The gain $AE$ achieved from the use of amplifiers is constant along white lines.%, where it equals the respective number in the white label.
		Interesting areas are those to the right of the second white line (at least two-fold gain from amplification) and to the left of the eighth black line counted from the right (indicating a spectral efficiency of at least $\SI{6}{bits/ s/ Hz}$).}\label{fig:savings-for-different-configurations}
\end{figure}

\begin{figure}[!hbt]
	\includegraphics[width=.49\textwidth,valign=t]{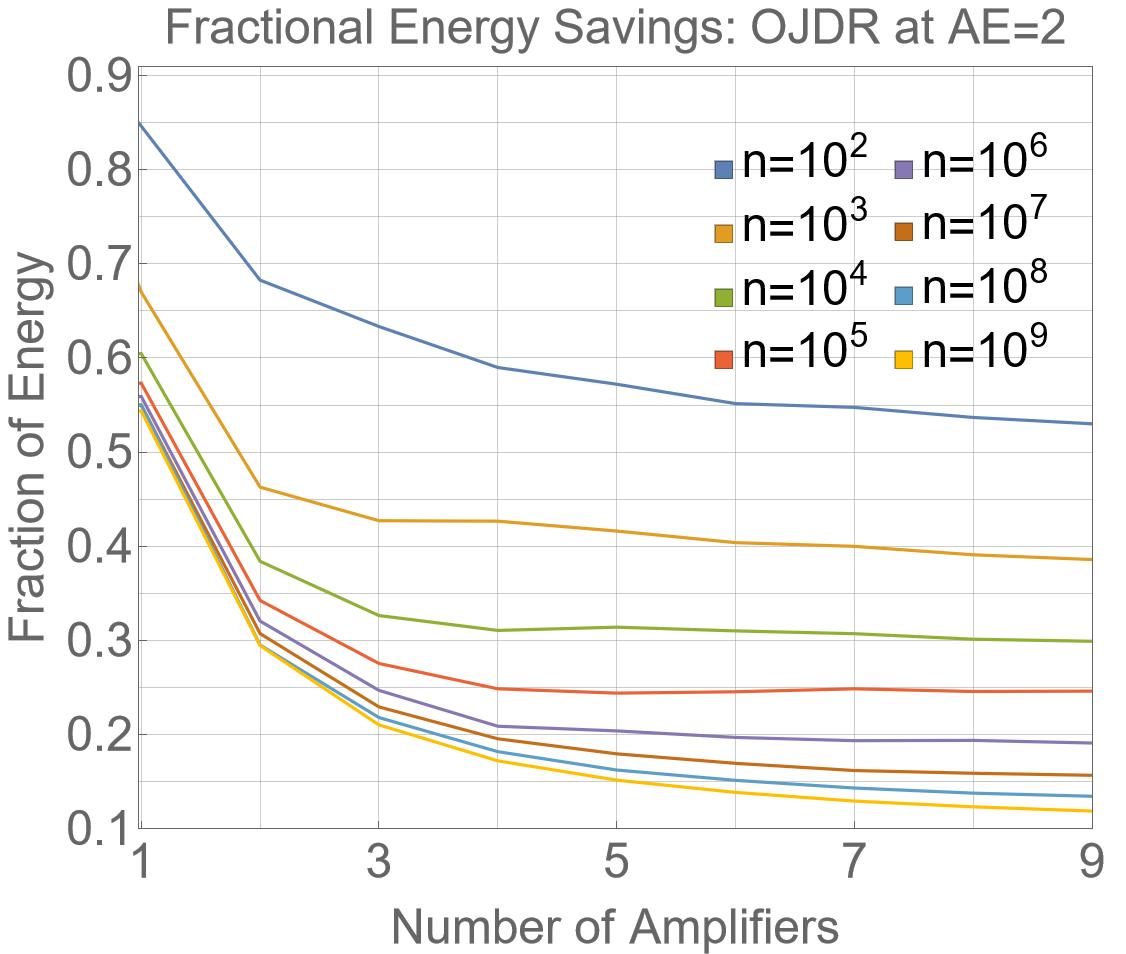}\hfill\includegraphics[width=.49\textwidth,valign=t]{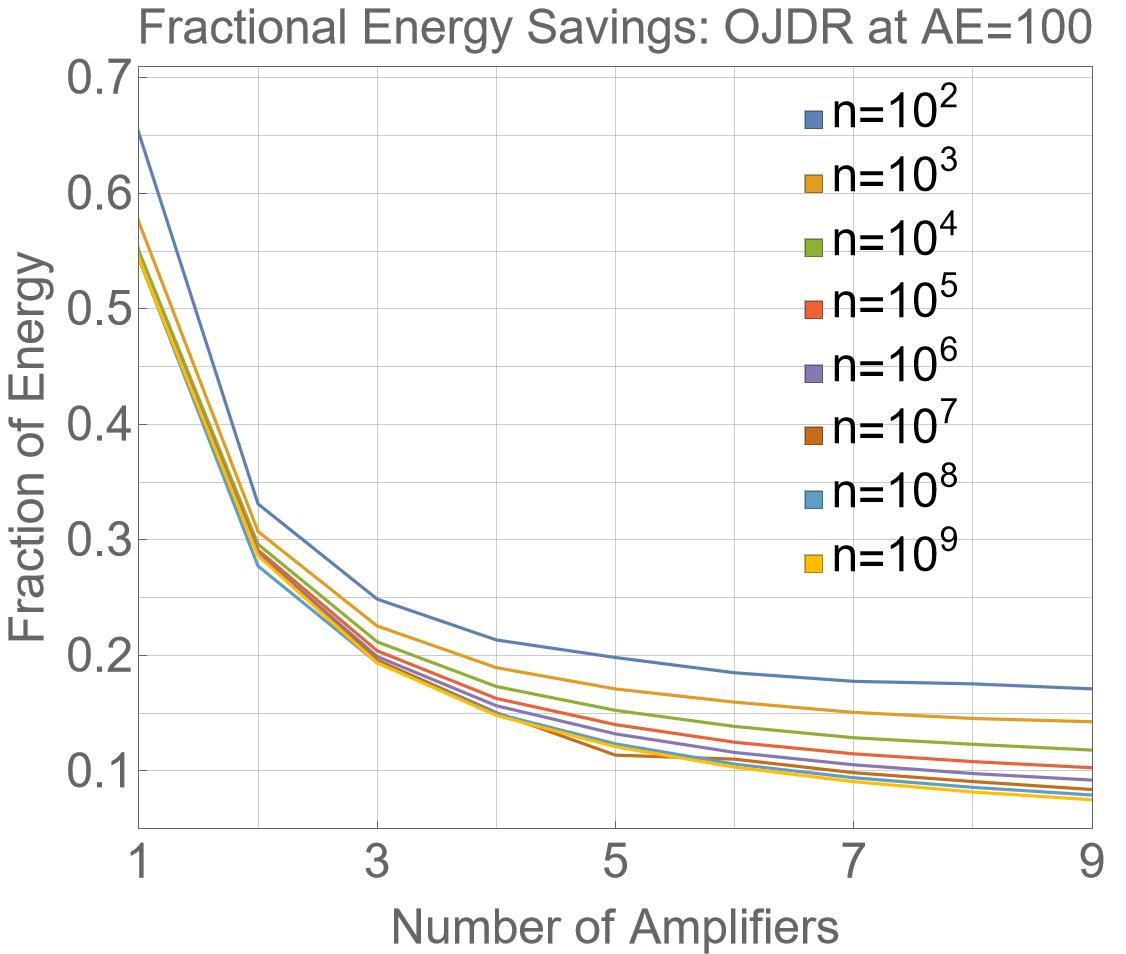}
	\caption{Displayed are the energy savings vs. the number of amplifiers for different values of the photon number $n$, at those points where $AE=2$ (left) and $AE=100$ (right). For the particular value of $n=10^7$, these are the points along the white lines labelled with $2$ and $100$ in Figure \ref{fig:savings-for-different-configurations}.} \label{fig:AE-overview}
\end{figure}

\begin{figure}[!hbt]
	\centering
	\includegraphics[width=\textwidth,valign=t]{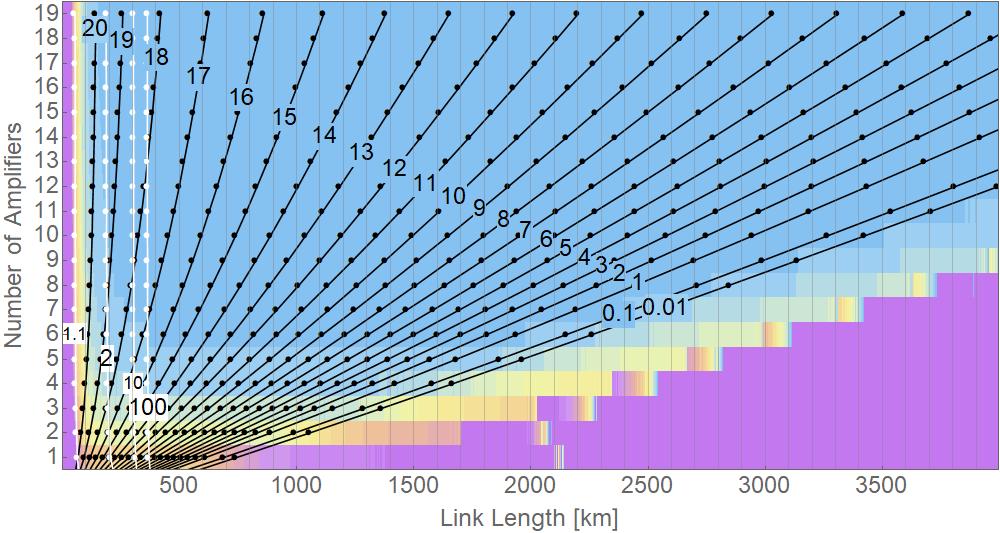}
	\includegraphics[width=\textwidth,valign=t]{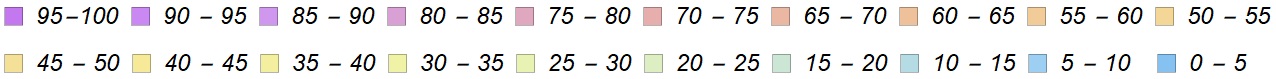}
	\caption{\label{fig:savings-for-different-configurations-overview}
		%\color{red}{Displayed are numerical approximations of the solution to problem REGS (left) and an overview over the behaviour of problem EGS for large distances and amplifier numbers (right), both at $n=10^7$.}
		Displayed is the percentage of energy used by the in-line amplifiers of links operating an OJDR relative to the same link with OSSR, as a function of link length and number of amplifiers.
		The photon number is set to $n=10^7$, SE is constant along black lines. The gain $AE$ achieved from the use of amplifiers is constant along white lines.}
\end{figure}

Using $\alpha$, $L$ and the number $K$ of segments as an input, we can now optimize the gain profile to solve problems \ref{eqn:mathematical-problem-statement1} and \ref{eqn:mathematical-problem-statement2} for concrete cases that are matched to the technical reality of optical fiber networks.
In Figure~\ref{fig:savings-for-different-configurations} we display an overview of our findings when the system is operating a single channel with typical values of wavelength $\lambda=\SI{1550}{nm}$, maximum input power $w=\SI{0.1}{mW}$ and attenuation coefficient $\alpha=\SI{0.05}{{km}^{-1}}$. Similar results for different system parameters are summarized in Figure \ref{fig:AE-overview}. An overview including more amplifiers and longer links is presented in Figure \ref{fig:savings-for-different-configurations-overview}. The algorithm employed to calculating the energy savings is explained in Methods \ref{appendix:algorithms}. Details regarding our choice of photon number per pulse are in Section \ref{subsubsec:parameter-choices}.

Figure \ref{fig:savings-for-different-configurations} shows that the use of OJDR technology allows for energy savings for all combinations of distance and number of amplifiers. We  plot the percentage of energy spent by OJDR with respect to a fully-amplified link with OSSR, i.e.,  $E_{\rm egs,regs}/E_{\rm sh}$, as a function of the link length and the number of segments. Furthermore, we consider two operational criteria to evaluate the practical relevance of our findings, aimed at identifying when the classical transmission system benefits the most from amplification:
\begin{enumerate}[(i)]
	\item the Shannon SE enhancement attainable by full amplification compared to no amplification, i.e. the ratio $AE:=\ssh^{\rm op}(n)/\ssh(n,\{1\}_{i=1}^K)$ (constant along white lines in Figure \ref{fig:savings-for-different-configurations}); 
	\item the Shannon SE achievable by full amplification $\ssh^{\rm op}(n)$ (constant along black lines in Figure \ref{fig:savings-for-different-configurations}).
\end{enumerate}

The record savings for a photon number of $n=10^7$ along the white line where $AE=2$ in Figure \ref{fig:savings-for-different-configurations} amount to $55\%$ and occur at a total link length of $\SI{225}{km}$ when the link is using one amplifier placed at a distance of $\SI{112.5}{km}$ from the sender. For this configuration, the SE reached with amplification is $\approx\SI{14.1}{bits/s/Hz}$ and that without amplification is $\approx\SI{7.0}{bits/s/Hz}$ when an OSSR is used and $\approx\SI{8.5}{bits/s/Hz}$ when an OJDR is employed.

A collective study over different photon numbers per pulse $n=10^2,10^3,\ldots,10^9$, modelling input energies $\SI{0.001}{mW},\ldots,\SI{10}{W}$ at the sender is depicted in Figure \ref{fig:AE-overview}.

We note further that even larger energy savings appear in situations of lesser practical relevance for commercial optical-fiber networks, which have been previously studied in terms of SE~\cite{Jarzyna2019a,Waseda10,Waseda11} and in terms of energy-efficiency in the limit where OSSR approximates OJDR ~\cite{Antonelli2014} (bottom-right and upper-left regions of Figure~\ref{fig:savings-for-different-configurations}). Indeed, at least $95\%$ of the energy can be saved by the OJDR in the regimes of long distance, though with extremely low achievable SE, and short distance, though with extremely low enhancement compared to the unamplified case.

Finally, the qualitative difference between our approximate solutions to problems EGS and REGS are shown in Figure \ref{fig:savings-for-different-configurations} to be quite small (on average over the data presented in Figure \ref{fig:savings-for-different-configurations} it amounts to $\approx5\%$, with a maximum difference in the energy savings of $\approx23\%$), implying that a simple decreasing-gain profile  already provides near-optimal energy savings for a large number of configurations.

\subsubsection{Parameter Choices}\label{subsubsec:parameter-choices}
The photon numbers per pulse used to calculate the Shannon and Holevo SE (\ref{eqn:shannon-S},\ref{eqn:holevo-S}) in this work are motivated from calculations targeted at fitting the model parameters to the real-world setting. 

We consider a system operating a single channel at the prototypical wavelength $\lambda=\SI{1550}{nm}$ of the C-band, and a laser output power of $w=\SI{0.1}{mW}$ Watts. The baud-rate, i.e. the number of pulses per second, is fixed to $b=\SI{80}{Giga\ pulses/s}$ (equivalent to $\SI{80}{Gbaud}$), a number which is in line with currently used numbers for fiber-optical transmission systems \cite{beyond100Gbaud}. 
%The historical C-band from $\SI{1530}{nm}$ to $\SI{1565}{nm}$ contained $80$ carriers with $\SI{50}{GHz}$ of guard band per carrier, leaving around $\SI{4,8}{GHz}$ per carrier. This results in a pulse duration of approximately $\SI{0.2}{ns}$. I took this information from here https://www.infinera.com/blog/all-you-ever-wanted-to-know-about-optical-transmission-bands/tag/optical/   -- another wihte paper by infinera is in /literature . In that white paper, they state the recent move to 100Gbaud.%
The energy per photon can be calculated using Planck's constant as 
\begin{align}
	e_p=\frac{h\cdot c}{\SI{1550}{nm}},
\end{align}
where $c$ is the speed of light. It satisfies $e_p\approx\SI{0.1}{AttJ}$. To be in a commercially interesting range of photon numbers we study situations where the laser output power $w$ in Watt is between $\SI{0,1}{\mu W}$ and $\SI{10}{mW}$. The laser output power fixes the number of photons per pulse via
\begin{align}
	w = e_p\cdot n\cdot b.
\end{align}
Correspondingly, we take a rounded value of $n=10^7$ to study transmission at $\SI{100}{mW}$. Our model predicts a large potential for energy savings resulting from the use of OJDRs. To make sure this prediction does not just depend on our choice of parameters, we study also systems with photon numbers $n\cdot10^{-3}$, $n\cdot10^{-2}$, $n\cdot10^{-1}$, $n\cdot10^{1}$ and $n\cdot10^{2}$. 

For the cases where the photon number per pulse equals $n$, $10n$ and $100n$,  non-amplified transmission over $\SI{80}{km}$ and with $\alpha=\SI{0.05}{1/km}$ our model predicts Shannon SEs of $\SI{17}{bits/s/Hz}$, $\SI{21}{bits/s/Hz}$ and $\SI{24}{bits/s/Hz}$. The data transmission rates corresponding to these parameter choices are then given by multiplying the Shannon SEs with the baudrate, and evaluate to $\SI{1.4}{Tbit/s}$, $\SI{1.7}{Tbit/s}$ and $\SI{1.9}{Tbit/s}$. A comparison with the literature reveals these numbers to be roughly in the range of the possible, yet unrealistically high. Since our model ignores any noise that is not resulting from optical amplification, and also disregards any effects arising from the nonlinear interaction of the pulses with the fiber, these deviations seem reasonable.

\section{Communication in the Quantum Limit}

We have so far quantified the possibility of energy savings in amplifiers based on the use of the OJDR, with a main focus on the current system parameters.

To point out the timeliness and usefulness of quantum receiver technology also with regards to future system architectures, we consider here in addition the limit of high baud rates. 
%As a primer, it is known from the literature and can also be inferred from Figure \ref{fig:AE-overview} that the advantage of the joint detection receiver in terms of spectral efficiency becomes more pronounced in the low photon-number limit. 

The key idea of this section follows the reasoning of operating a communication system in a parameter regime which is chosen in an attempt to harness the superior performance of quantum data transmission techniques. When a power constraint is imposed on a communication link with a large enough spectral bandwidth, high-baud-rate transmission systems naturally operate on a low photon number per pulse. The technological feasibility of such a system must then be answered based on the imposed bandwidth limitations and the available processing speed of the receiver in conjunction with the expected added value arising from the novel system design. 

The interesting observation one can make based on formulas \eqref{eqn:shannon-S} and \eqref{eqn:holevo-S} is that for a fixed maximum allowed number $N$ of photons per second, the number of photons per pulse (at the transmitter) satisfies $n=N/b$ where $b$ is the number of pulses per second (also called baud-rate). At a given attenuation $\tau$ and noise level $\nu$ per pulse, the number of bits per second that can be transmitted using an OSSR saturates for high baud-rates:
\begin{align}
	\lim_{b\to\infty}\log\left(1+\frac{\tau\cdot \tfrac{N}{b}}{1+\nu}\right)\cdot b=\frac{N\cdot\tau}{(1+\nu)\ln(2)}\leq \frac{N\cdot\tau}{\ln(2)},
\end{align}
whereas the corresponding quantity for an OJDR diverges at zero thermal noise:
\begin{align}
	\lim_{b\to\infty}\left(g\left(\tau\cdot \tfrac{N}{b}+\nu\right) - g(\nu)\right)\cdot b=N\cdot\tau\cdot\log\left(1+\tfrac{1}{\nu}\right)
\end{align}
Thus since the spectral bandwidth scales with baud rate $b$, settings of low thermal noise and %(LOW?) 
high spectral bandwidth will allow the OJDR to outperform the OSSR by any %(FIXED? nu-dependent) 
arbitrary factor. 

If also the thermal noise per second stays constant %(CAREFUL how we motivate this)
while the baud-rate increases, this noise per pulse equals $\nu/b$. In such a scenario, the limiting expression for the OJDR is obtained by noting that for any $x,y>0$ we have
\begin{align}
	g(\tfrac{x}{b})\cdot b = x\cdot\log(1+\tfrac{b}{x})+b\cdot\log(1+\tfrac{x}{b}).
\end{align}
Using the limiting expressions $\lim_{b\to\infty}b\cdot\log(1+x/b)=x/\ln(2)$ and $\lim_{b\to\infty}\log(1+\tfrac{b}{x})-\log(1+\tfrac{b}{y}) = \log(\tfrac{y}{x})$ we know that for each choice of $(N,\tau,\nu)$ there exists a sequence $(\epsilon_b)_{b\in\mathbb R}$ and a constant $\epsilon=\epsilon(N,\tau,\nu)$ such that $\epsilon_b\to\epsilon$ and
\begin{align}
	\left(g\left(\tau\cdot \tfrac{N}{b}+\tfrac{\nu}{b}\right) - g(\tfrac{\nu}{b})\right)\cdot b=\tau\cdot N\cdot\log(\tfrac{b}{\tau\cdot N+\nu})+\epsilon_b.
\end{align}
%(I get log(b/(\tau N + \nu)) only, not a big difference anyway)
%I'll try to spell it out here to get the clarity:
% g(x/b)b
%   = (x + b)log(1 + x/b) - x log(x/b)
%   = (x + b)log(1 + x/b) - x log(x/b)
% (g((N+V)/b) - g(V/b))/b 
%   = (N + V + b)log(1 + (N+V)/b) - (N+V)log((N+V)/b)
%       - (V+b)log(1+V/b) + Vlog(V/b)
%   = ... + Nlog(b/(N+N))
% (you are right)

Thus for every choice of $N,\tau$ and $\nu$ the dependence of the channel capacity on the baud-rate is effectively logarithmic and grows unbounded as $\tau\cdot N\cdot\log(\tfrac{b}{\tau\cdot N+\nu})$ when using an OJDR. In sharp contrast, the capacity of the same channel using an OSSR is limited by $\frac{\tau\cdot N}{\ln(2)}$. As the latter bound applies in particular also when amplifiers are used, the use of high baud-rates provides an opportunity for non-amplified networks. This new possibility is of high relevance for future integrated classical- and quantum networks, which might need to transmit classical messages on the same shared medium as quantum services. For transmission and maintenance of entangled states, noise is a challenging problem. Thus for integrated networks, where noise needs to be avoided at all costs, it is important to observe that there exist quantum data transmission methods which avoid the noise induced by amplification in data transmission, so that the resulting noiseless networks become an ideal starting point for quantum network development.

To investigate the current hypothetical advantage of quantum communication methods in high baud-rate transmission, it is insightful to consider parameters matched to existing networks. %Setting aside practical considerations outside the range of theoretical tools used in this work, wide-band %(aren't we doing NARROWBAND here? The baud-rate you mention here probably includes parallel pulses at different frequencies so you might need to correct it) 
Transmission with center wavelength at $\SI{157.5}{nm}$ would allow for a maximum baud rate in the order of $b\approx\SI{18000}{GigaBaud}$. At this baud rate, the number of bits per second for transmission over a cable of length 
$L=\SI{185}{km}$ at 
$\SI{1}{mW}$ is 
$\approx\SI{1}{TeraBit/s}$ with an OSSR and 
$\approx\SI{4.3}{TeraBit/s}$ with an OJDR - more than a $4$-fold increase in data transmission rate, which is mainly limited by the carrier frequency and the spectral bandwidth limitations imposed by the physical properties of optical fiber in the C-band, plus potentially the processing speed of electronic gates in the receiver.
\begin{figure}[!hbt]
	\begin{center}
		\includegraphics[width=\textwidth]{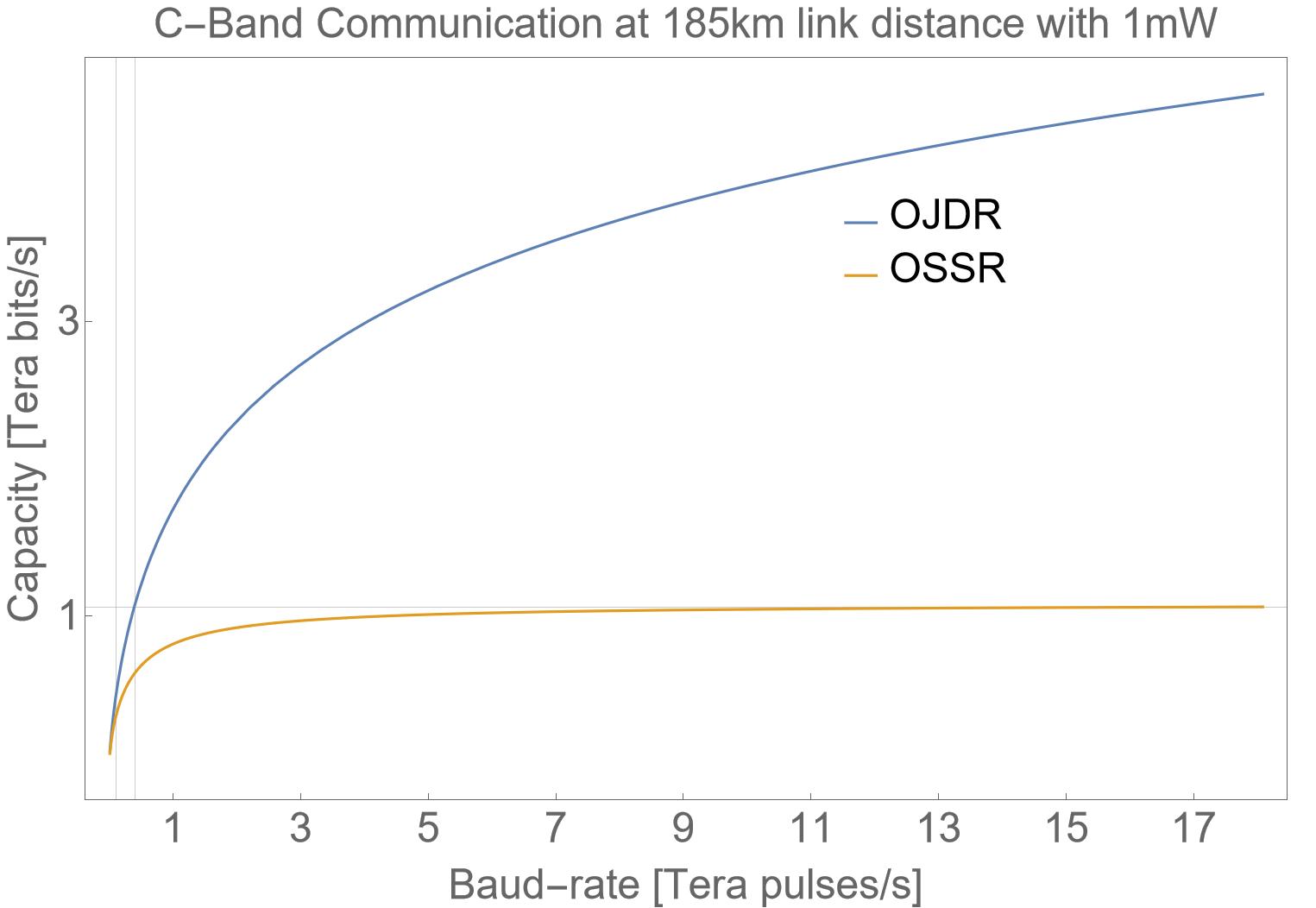}
		\caption{Displayed is the growth of capacity in bits per second as a function of baud-rate at a distance of 185km with 1mW transmit power at the sender and attenuation coefficient of $\alpha=\SI{0.05}{km^{-1}}$. The vertical line at baud-rate $b=0.1\cdot 10^{12}$ shows where a current commercial system operates. The vertical line at $0.44\cdot10^{12}$ marks the beginning of the quantum limit.}
	\end{center}
\end{figure}
For both problems, solutions are in principle possible: hollow-core fibers can be better tailored to the transmission of ultra-short pulses \cite{broadband-hollow-core,broadband-hollow-core-communications,hollowCore700nmWindow}, and optical processing as a physical-layer technique can utilize the principles of quantum-mechanical detection based on newly developed integrated optics techniques.

Spreading the signal energy for data transmission over a wide spectral range might in addition open up a way for co-existence of data transmission with other services of future quantum networks. Our analysis thus suggests an exciting new venue for quantum communication network development.

Finally we observe that, though the realization of a full-fledged OJDR is still an open problem, the Hadamard receiver~\cite{Guha11,Guha2012,Rosati16c,Klimek2015}, realizable with integrated photonics, offers the opportunity to observe the high-baud-rate advantage enabled by a quantum receiver in a realistic setting. In the above setting of communicating in the C-band over $\SI{185}{km}$ at $\SI{1}{mW}$, Hadamard codes with orders $4, 8, 16, 32$ would offer respective transmission rates of $\SI{1.3799}{Tbit/s}$, $\SI{1.91191}{Tbit/s}$, $\SI{2.18987}{Tbit/s}$, $\SI{2.0744}{Tbit/s}$. These rates can be calculated in the zero-noise case as
\begin{align}
	\frac bk \left(1-\exp\left(-\frac{k\tau n}{b}\right)\right)\log k,
\end{align}
using for example Ref.~\cite{Guha11}.

\section{Discussion}
We have shown that the introduction of a quantum detection method, the OJDR, can reduce the energy cost of a fiber-optical communication line by up to $\approx55\%$, in a practically relevant parameter regime where optical amplifiers enable to maintain a large communication rate. 

This determines, for the first time to our knowledge, a striking energy advantage of quantum vs. classical detection methods for the transmission of classical information on optical fiber. Our results highlight the relevance of QIP in communication beyond the traditional settings of secure-communication and entanglement-transmission, opening up a third research direction with potentially closer-term applications, which is fully compliant with the vision of all-optical networking \cite{all-optical-networking-evolution}.

It is well-known that the performance of quantum receivers approaches that of their classical counterparts under high noise. Resulting from this fact is a perception that quantum receiver technology might not be relevant in practical applications, where noise cannot be avoided. Our approach inverts the argument: by reducing the amplifier gains along the line we reduce the noise resulting from spontaneous emission, thereby moving towards the regime where the quantum receiver is superior while at the same time reducing the energy consumption of the amplifiers.

Along this line of argument, we have shown that even a complete removal of amplifiers is possible, given enough spectral bandwidth supporting high baud-rates. We have described the respective scaling laws of the data transmission rates with the baud-rate for both OJDR and OSSR technology.

We stress that, although our analysis can be refined by employing more complex models of energy consumption, including non-linearities and the energy transfer towards the amplifiers, these effects can only increase energy consumption and hence provide further arguments in favour of the OJDR. %Furthermore, it might be interesting to study the robustness of our results to other realistic noise effects~\cite{Fanizza2020b,Cacioppo2021a}.
Our results renew the urgency of devising an explicit optical receiver design that can approximate the OJDR at all signal energies and set the stage for a redesign of established data transmission technology based on a use of quantum-mechanical principles.

%Throughout, we make the idealized assumption that the energy consumption of the amplifiers deployed on the link scales linear with the gain. Also, we do not consider the problem of energy transfer towards the amplifiers, which could provide further arguments in the favor of a use of OJDR technology.  
%Finally, we stress that  (see Appendix...).  
%While notable differences between the algorithms are visible, brute-force optimization at selected configurations of link length and number of amplifiers did not change the overall picture, which is summarized as follows: For every length $L$ of the communication link, and for every number $K$ of segments, using quantum communication technology at the receiver results in energy savings for the amplifiers without any sacrifice in terms of spectral efficiency.
%Further, we observed the possibility to harvest energy savings up to a record of $57.5\%$ by deploying  an OJDR in a situation where the parameters of the corresponding classical system using conventional decoding technology are:\\
%Spectral efficiency above $8.6\ \mathrm{bits/s/Hz}$ with the amplifier operating at maximum gain, spectral efficiency $4.3\ \mathrm{bits/second/Hertz}$ with the amplifier turned off, $L=148.5\ \mathrm{km}$, $\alpha=0.05\ \mathrm{km}^{-1}$, and $n_S=31211.5$. 
%These numbers mimic the context of communication in the C-band at a the telekom wavelength of $1550\mathrm{nm}$ and a baudrate of $100\mathrm{Gbaud}$.

In what follows, we list implications of our results which are relevant to other research domains in networking. Namely, OJDR technology can:\\
{\bf1) provide advantages already today:} these come in terms of energy savings at amplifiers.\\
{\bf2) increase the length of non-amplified links:} If new types of optical fiber with more spectral bandwidth were deployed, the observed initial savings could eventually translate into a complete removal of the amplifiers, without any sacrifice on the data rates.\\ 
{\bf3) lay the ground for quantum network development:}
A removal of amplifiers decreases the overall noise in the channel, and we can expect this to imply an increase of the quantum-data-transmission capacity. For example, consider the most basic instance of our communication line (Fig.\ref{fig:comm-line}) with $K=2$ optical-fiber links of loss $\eta$, separated by a quantum-limited amplifier of gain $G$. The overall channel, determined by \eqref{eqn:definition-of-tau-and-nu}, can be described as a thermal-attenuator bosonic channel with attenuation coefficient $G\tau^2$ and extra-noise coefficient $\frac{\tau(\tau+2)(G-1)}{2(1-G\tau^2)}$, in the notation of \cite[Eq. (11)]{Rosati2018a}. Both the upper and lower bounds~\cite[Eqs. (18,40)]{Rosati2018a} appear to be decreasing as a function of the gain and hence one can expect that quantum communication benefits from the complete removal of amplifiers along the line\footnote{The lower bound was first found in \cite{quantumCapacityOfBosonicChannel}. Better upper bounds at the considered noise and attenuation levels have been obtained in \cite{Fanizza2021}. Obviously, the lack of an exact formula makes the decrease of the quantum capacity as a function of the gain a well-motivated conjecture, rather than a certainty. Furthermore, a more careful analysis of the above statement should rely on the study of the joint capacity region for classical and quantum communication~\cite{Wilde2012b}.}.
\\
{\bf4) simplify network deployment and maintenance:} The use of this technology in conjunction with high baud-rates offers a possibility to simplify deployment and operation of data networks that can be operated without amplifiers.\\
{\bf5) offer a way to evade noise induced by the Kerr effect:}
A consequence of our work is the emergence of transmission in the quantum limit as a new design option for fiber-optic networks. For such networks, the non-linear Shannon limit resulting among others from the Kerr effect imposes severe practical boundary conditions  \cite{capacityLimitsOfOpticalFiberNetworks}. Optimization under these boundary conditions is an active research field \cite{shapingLightwaves}. The Kerr effect itself induces a specific type of noise to the system which depends on the signal energy \cite{nonlinearLimits} and thus eventually prevents a growth of the system capacity with the signal energy. Our approach does instead advertise communication in a low-energy regime where the Kerr effect has little impact, and where the capacity loss incurred from the lower energy is compensated for by high baud-rates. This approach is in line with the historic trend of pushing baud-rates to higher and even higher numbers, which is documented e.g. in \cite{baudRateTrend}.
%https://www.gazettabyte.com/home/2022/2/15/building-the-data-rate-out-of-smaller-baud-rates.htmlXXXXX
\\\\
{\bf The following is an important limitation in our analysis:}   
We point out that our analysis rests on the established formulas for Shannon- \cite{Shannon1948b} and Holevo \cite{Holevo1973} capacity, which take into account only attenuation and thermal noise. Any further effects arising from a use of ultra-short pulses are beyond the scope of our analysis. Thus our work justifies a more in-depth analysis of transmission media when operated in the quantum limit, which we define as the baud-rate at which the data transmission capacity of the link operated with an OJDR exceeds the upper bound imposed on that same link when operated with an OSSR.

In summary, our results mark the importance of QIP in terms of energy-efficiency, which is a metric of growing impact for optical network engineering. They vindicate the central role of quantum detection in optical communication technologies, and motivate a re-examination of quantum advantage beyond its traditional meaning in computation, metrology and communication.

\section{Methods}

\subsection{Shannon and Holevo SEGS}\label{appendix:op-gains}
%Here we prove several properties related with the optimal SE's. 
For a given segment $i$ in the transmission line, let us define the cumulative attenuation and noise coefficients of the future segments as
\begin{equation}
	\tau_{>i}:=\eta^{K-i}\prod_{j=i+1}^K G_j,\quad \nu_{>i}:=\sum_{j=i+1}^K (G_j-1)\cdot \tau_{>j}.
\end{equation}
In terms of these, the total attenuation and noise coefficients can then be written as
\begin{equation}\label{eqn:coeff-deco}
	\tau=\tau_{>i}\cdot\tau_i,\quad \nu=\tau_{>i}\cdot\nu_i + \nu_{>i}.
\end{equation}
\newline

\subsubsection{Non-increasing gains maximize the SE}\label{appendix:decreasing-gains-optimal}
We start by proving that increasing gains are sub-optimal for information transmission on our multi-span amplified optical-fiber transmission line. 
Consider a channel where two consecutive gains are in increasing order, i.e., $G_i<G_{i+1}$. Then, the channel obtained by switching such gains, i.e., $G_i':=G_{i+1}>G_{i+1}':=G_i$, is characterized by the same attenuation coefficient $\tau'=\tau$, since it contains the product of all the gains, and a smaller noise coefficient $\nu'<\nu$. Indeed, we can write the new noise coefficient as
\begin{align}
	\nu'&=\tau_{>i}'\cdot\nu_i' + \nu_{>i}'\nonumber\\
	&=\tau_{>i+1}\cdot G_{i+1}'\eta\cdot( G_i'\eta\cdot\nu_{i-1}+G_i'-1)+ (G_{i+1}'-1)\cdot\tau_{>i+1} + \nu_{>i+1},
\end{align}
where we used the fact that all gains at segments different from $i$, $i+1$ remained unchanged. 
Using the same decomposition for $\nu$, and taking into account the definition of $G'$, we can write the noise-difference as
\begin{align}
	\nu-\nu'&=\tau_{>i+1}\cdot[-\eta\cdot(G_{i+1}-G_{i})+G_{i+1}-G_i]\nonumber\\
	&=\tau_{>i+1}\cdot(G_{i+1}-G_i)\cdot(1-\eta)>0.
\end{align}
Hence, the channel with increasing gains can be obtained from the channel with decreasing gains by composition with a Gaussian additive-noise channel of noise parameter $\nu-\nu'$. By the data-processing inequality applied to the classical or quantum mutual information, this implies that the channel with decreasing gains can transfer more information, independently of the communication method employed~\cite{WildeBOOK}. 
\newline

\subsubsection{Shannon SEGS}\label{appendix:shannon-segs}
Let us now prove that the Shannon SE is monotonically increasing as a function of the gains $G_{i<K}$. First observe that $\ssh$ is an increasing function of the signal-to-noise ratio $\snr:=\tau\cdot n/(1+\nu)$. Then express the derivatives of the attenuation and noise coefficients with respect to gain $G_{i}$ as follows:
\begin{align}
	\frac{\partial \tau}{\partial G_i} = \tau_{>i}\cdot\eta=\frac{\tau}{G_i},\quad
	\frac{\partial \nu}{\partial G_i} = \tau_{>i}\cdot(\eta\cdot\nu_{i-1}+1) = \frac{\nu-\nu_{>i}+\tau_{>i}}{G_i},
\end{align}
where we have employed \eqref{eqn:coeff-deco}.
Substituting these expressions into the derivative of the $\snr$ we finally obtain
\begin{align}
	\frac{\partial \snr}{\partial G_i}&=\frac{\snr}{G_i}-\frac{\snr}{1+\nu}\cdot \frac{\nu-\nu_{>i}+\tau_{>i}}{G_i}\nonumber\\
	&=\frac{\snr}{G_i}\cdot\frac{1-\tau_{>i}+\nu_{>i}}{1+\nu}\geq0,\label{eq:derSnr1}
\end{align}
since $\tau_i\leq1$ by \eqref{eqn:power-constraint}. In particular, equality is attained only for $i=K$, while for $i<K$ the Shannon SE is strictly increasing and hence the optimal gain profile is that with $G_{i<K}=G_{\max}$.
\newline 

\subsubsection{Holevo SEGS}\label{appendix:holevo-segs}
The results are quite different for the Holevo SE. Considering that $g'(x)=\log(1+1/x)$, we can write
\begin{equation}\label{eqn:der-hol}
	\begin{aligned}
		\frac{\partial \sho}{\partial G_i}&=\left(\frac{\tau}{G_i}\cdot n+\frac{\nu-\nu_{>i}+\tau_{>i}}{G_i}\right)\log\left(1+\frac{1}{\tau\bar n+\nu}\right)    \\
		&-\left(\frac{\nu-\nu_{>i}+\tau_{>i}}{G_i}\right)\log\left(1+\frac{1}{\nu}\right)   \\
		&=\frac{1}{G_i}\log\frac{f_{\beta_i}(\tau\cdot  n+\nu)}{f_{\beta_i}(\nu)},
	\end{aligned}
\end{equation}
where $f_\beta(x):=(1+\frac1x)^{x+\beta}$ and $\beta_i=\tau_{>i}-\nu_{>i}\leq1$. 
Hence, the sign of \eqref{eqn:der-hol} is determined by the behaviour of $f$ for different values of $\beta$. It can be checked that $f_\beta(x)$ is a monotonically increasing function of $x$ for $\beta\leq0$, while it is monotonically decreasing for $\beta\geq1/2$; instead, for $0<\beta<\frac12$ it has a global minimum at intermediate values of $x$, whose position moves from $0$ to $\infty$ as $\beta$ increases. 
We conclude that, as a function of $G_i$, the Holevo SE can be found in three distinct regimes: 
\begin{enumerate}[(i)]
	\item  no-amplification regime: in this case it holds $\beta_i\geq1/2$, hence \eqref{eqn:der-hol} is strictly negative for all $n>0$, independently of the other gains, and $\sho$ is maximum at $G_i=1$; 
	\item maximum-gain regime: in this case it holds $\beta_i\leq0$, hence \eqref{eqn:der-hol} is strictly positive for all $n>0$, independently of the other gains, and $\sho$ is maximum at the maximum allowed value of gain $G_i=G_{\max}$; 
	\item intermediate regime: in this case it holds $0<\beta_i<1/2$ and the sign of \eqref{eqn:der-hol} must be determined on a case-to-case basis depending on $\tau$, $\nu$, $n$ and the optimization cannot be carried out independently of the other gains. Still, it can happen that the optimal gain value is minimum. Indeed, calling $x_{\beta_i}$ the minimum of $f_{\beta_i}(x)$, suppose that all the gains previous to $G_i$ are at their maximum value and $\tau\cdot n+\nu<x_{\beta_i}$; then \eqref{eqn:der-hol} is strictly negative for that particular choice of gains but also for any smaller values of those gains; since decreasing any gain will decrease both $\tau$ and $\nu$, preserving the previous inequality. Hence it will be possible to set $G_i=1$ also in this case.
\end{enumerate}
Furthermore, another property that might be helpful in the optimization is that $\beta_i$ is a non-decreasing sequence, since
\begin{align}
	\beta_i&=\tau_{>i}-(G_{i+1}-1)\cdot\tau_{>i+1}-\nu_{>i+1}\\
	&=\tau_{>i}(1-\frac1\eta)+\beta_{>i+1},
\end{align}
where we have written explicitly the first term in the decomposition of $\nu_{>i}$ and recalled that $\tau_{>i}=\eta\cdot G_i\cdot\tau_{i-1}$.

In conclusion, the analytical optimization of $\sho$ seems difficult. Still, the above-described properties allow very easily to identify parameter regimes that ensure that at least one gain is minimum, e.g., the condition $1/2\leq\beta_{K-1}=\eta$ implies $G_{K-1}^{\rm op}=1$, proving that $\sho^{\rm op}>\sho(n,\{G_{i<K}=G_{\max},G_K=1\})$ in general.

The nontrivial dependence of Holevo SE on the gain profile can be illustrated in the situation of continuous amplification:

\subsection{Energy savings with continuous amplification}\label{appendix:cont-amp}
Here we show that energy gains can be obtained by using an OJDR vs. an OSSR even in the continuous-amplification limit. This is in contrast to the claim of Ref.~\cite{Antonelli2014}, where the Holevo SE was studied only in the setting where an OJDR is comparable with an OSSR. 

The optimal Shannon SE in the continuous-amplification limit can be determined by taking $K\rightarrow\infty$ in \eqref{eqn:shannon-op} and observing that $\eta\approxeq1-\alpha L/K$ in this limit:
\begin{equation}
	\ssh^{{\rm op},\infty}(n)=\log\left(1+\frac{e^{-\frac{\alpha L}{n+1}}\cdot n}{1+\left(1-e^{-\frac{\alpha L}{n+1}}\right)\cdot n}\right).
\end{equation}
The corresponding energy cost is simply
\begin{equation}
	E_{\rm sh}^{\infty}(n)=\alpha L\cdot n.
\end{equation}

In the case of the Holevo SE, we start by considering $K$ finite segments, out of which only the first $\gamma\cdot K$ are amplified, for $\gamma\in[0,1]$. This is a sub-optimal strategy for Holevo SEGS that still captures most of the relevant features of the problem and that can actually be optimal in some cases, based on the analysis of Methods~\ref{appendix:holevo-segs}.
The corresponding attenuation and noise coefficients can be written as
\begin{equation}
	\tau_K(\gamma):=e^{-\alpha L}\cdot\left(\frac{1+n}{1+\eta n}\right)^{\gamma K}, \quad \nu_K(\gamma)=\left(e^{-\alpha L(1-\gamma)}-\tau_K(\gamma)\right)\cdot n,
\end{equation}
and, in the limit $K\rightarrow\infty$ with $\gamma$ fixed approach
\begin{equation}
	\tau_\infty(\gamma)=\exp\left[-\alpha L\left(1-\frac{\gamma\cdot n}{1+n}\right)\right], \quad \nu_\infty(\gamma)=\left(e^{-\alpha L(1-\gamma)}-\tau_\infty(\gamma)\right)\cdot n.
\end{equation}
The corresponding SE and energy cost are
\begin{align}
	&\sho^{\infty}(n,\gamma)=g\left(e^{-\alpha L(1-\gamma)}\cdot n\right)-g\left(\nu_\infty(\gamma)\right)\\
	&E_{\rm ho}^\infty(n,\gamma)=\alpha L \gamma\cdot n.
\end{align}

\begin{figure}[!t]
	\centering
	\includegraphics[width=.49\textwidth]{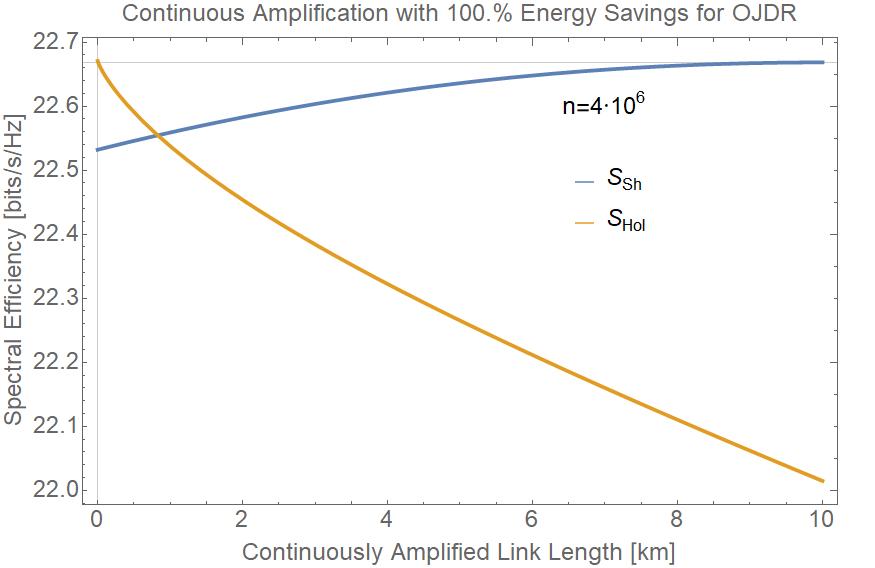}
	\includegraphics[width=.49\textwidth]{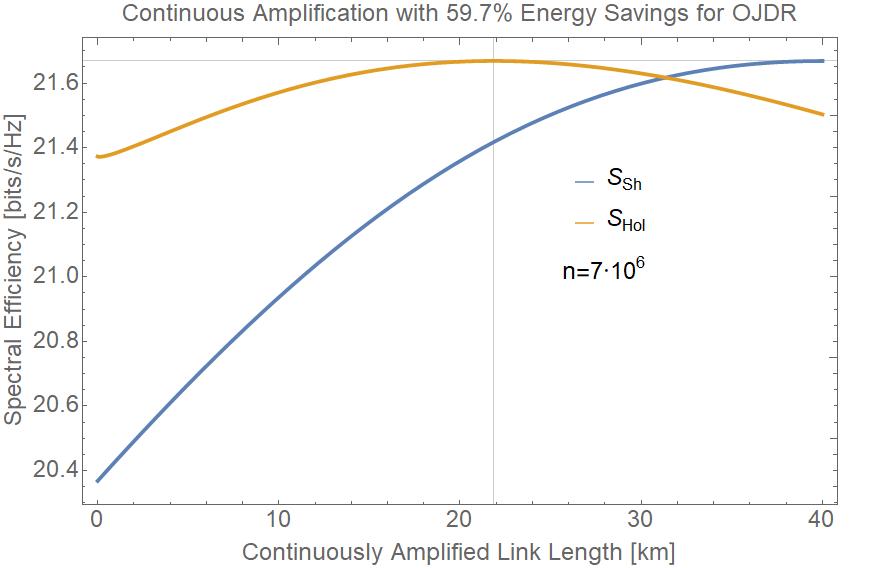}\\\ \\
	\includegraphics[width=.49\textwidth]{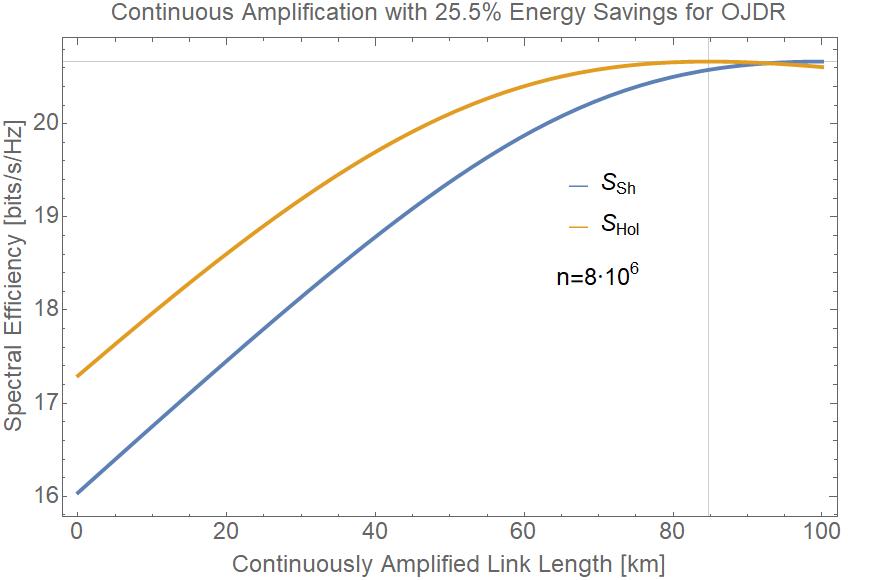}
	\includegraphics[width=.49\textwidth]{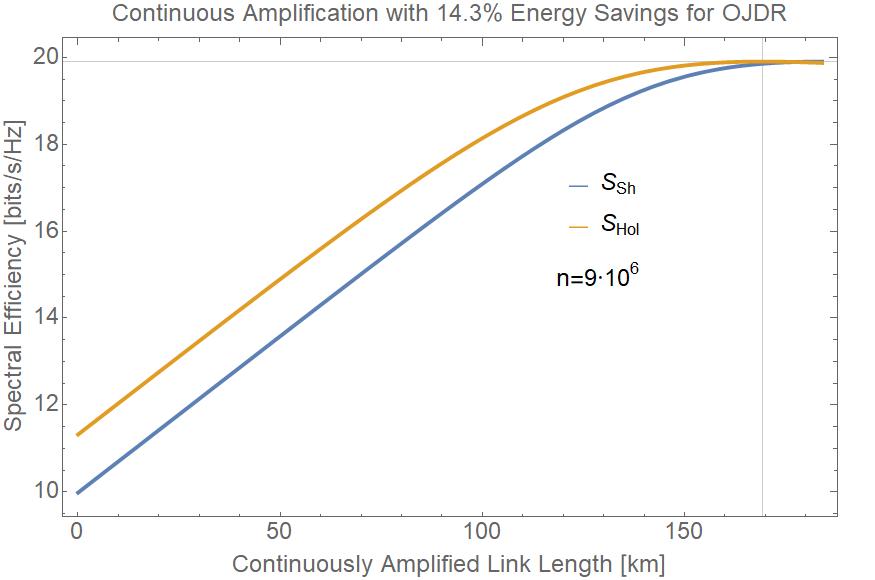}
	\caption{Spectral efficiency of communication links with $L=\SI{10}{km}$, $\SI{40}{km}$, $\SI{100}{km}$, $\SI{184}{km}$ (counting top down and from left to right) and $\alpha=0.05\mathrm{km}^{-1}$. The $x$ axis denotes the length $L_1$ of the initial segment that is fully amplified. As expected, the Shannon spectral efficiency is monotone in $L_1$. However, the spectral efficiency when using an OJDR is at the same value as the one when using an SSR at lengths $L_1<L$ in all depicted scenarios. Moreover, the power constraint on the link can be reduced when OJDR is used so that energy savings can be harvested.}
	\label{fig:continuous-amplification-limit}
\end{figure}
It is now possible to define an optimization problem similar to EGS for this on-off strategy:
\begin{equation}\begin{aligned}\label{eqn:mathematical-problem-statement-onOff}
		E_{\rm oo}:=\underset{n,\gamma}{\mathrm{minimize}}\ & E_{\rm ho}^\infty(n,\gamma)\\
		\mathrm{subject\ to:}\ &0\leq n\leq n_0, 1\leq \gamma\leq 1\\  &\sho^{\infty}(n,\gamma) \geq \ssh^{\rm op,\infty}(n_0).
\end{aligned}\end{equation}
Here, both the maximum power at any point on the segment can be optimized and the length of the initial, amplified segment. Setting e.g. the power constraint for communication with the OSSR to $n_0=10^7$, a multitude of different energy savings can be realized, which highlight the nontrivial dependency of the communication link on the gain profile even in the limit of continuous amplification. Note that all situations depicted in Figure \ref{fig:continuous-amplification-limit} have a value of $AE<2$, except the one with $L=\SI{184}{km}$ where $AE\approx2$ holds.

%Even with such a simple strategy it becomes clear that the problem of optimal continuous amplification in the sense of the Holevo-SEGS is quite nontrivial, as the optimal selection of 

%The case displayed in Figure \ref{fig:continuous-amplification-limit} shows that using  an OJDR can be superior even to continuous amplification. However, the downside of this example is that even continuous amplification increases the spectral efficiency by less than $10\%$. 

%It is however important to point out that it is optimal to amplify only the first roughly $17\mathrm{km}$ of the total link when using  an OJDR, since otherwise the spectral efficiency decays again. Thus computing the quantum limit of the communication link under a fixed constraint on the total energy at every point along the link is nontrivial. In the particular case treated here one realizes a reduction in energy consumption from amplification by more than $50\%$. 

%\input{paper/structure-of-feasible-region}

\subsection{Algorithm}\label{appendix:algorithms}

Here we describe the algorithm we employed to solve the optimization problems EGS and REGS. 

For both these problems, since the feasible region is closed and bounded and the objective function continuous, the Weierstrass theorem guarantees the existence of a global optimum. However, the function $\sho$ is in general neither convex nor concave. 

We applied several numerical methods to find approximate solutions to problems \ref{eqn:mathematical-problem-statement1} and  \ref{eqn:mathematical-problem-statement2}. The most efficient one turned out to be a gradient-based method that we describe in the following. The general approach is identical for both EGS and REGS and also for the optimization of energy consumption when homodyne detection is employed.

Starting from maximum gains, the method iterates between optimizing the energy consumption $E$ (or $E'$ for REGS) following the gradient $\nabla E$ (or $\nabla E'$ for REGS), and then optimizing energy consumption along a surface where $\sho$ is constant. Searching the solutions to REGS is particularly fast since $\nabla E'$ is constant. 

Assuming that $G_K=1$ is obeyed throughout so that it does not need to be listed as a variable, the two sub-routines are to follow the gradient of $E$ with decreasing speed for several steps as described in Algorithm \ref{algo:follow-E-gradient}, and to follow along a surface where $\sho$ is constant as described in Algorithm \ref{algo:walk-on-surface}.
\begin{algorithm}
	\caption{energyGradient}\label{algo:follow-E-gradient}
	\begin{algorithmic}[1]
		\Require $G_1,\ldots,G_{K-1},\ a,\ n,\ SE$
		\State $\mathrm{s} = 1$
		\State $\mathrm{stepSize}\Leftarrow (G_{\max}-1)/2^{-s}$
		\While{$s<a$}
		\State $x\Leftarrow\|\nabla E\|^{-1}\cdot \nabla E$
		\State $(\hat G_1,\ldots,\hat G_{K-1})\Leftarrow (G_1 -x_1\cdot \mathrm{stepSize},\ldots,G_{K-1}-x_{K-1}\cdot \mathrm{stepSize})$
		\If{$\sho(n,\{\hat G_i^k\}_{i=1}^{K})>SE$ and $G_{\max}\geq\hat G_i\geq1\forall\ i\in\{1,\ldots,K-1\}$}
		\State $(G_1,\ldots,G_{K-1})\Leftarrow(\hat G_1,\ldots,\hat G_{K-1})$
		\Else{ $s\Leftarrow s+1$}
		\State $\mathrm{stepSize}\Leftarrow (G_{\max}-1)/2^{-s}$
		\EndIf
		\EndWhile
		\State \Return $G^K$
	\end{algorithmic}
\end{algorithm}

\begin{algorithm}
	\caption{spectralSurface}\label{algo:walk-on-surface}
	\begin{algorithmic}[1]
		\Require $G_1,\ldots,G_{K-1},\ a,\ n,\ SE$
		\State $\mathrm{s} = 1$
		\State $\mathrm{stepSize}\Leftarrow (G_{\max}-1)/2^{-s}$
		\While{$s<a$}
		\State $\mathrm{eG}\Leftarrow\|\nabla E\|^{-1}\cdot \nabla E$
		\State $\mathrm{sG}\Leftarrow\|\nabla\sho\|^{-1}\cdot\nabla \sho$
		\State $x=eG-\langle sG,eG\rangle\cdot sG$
		\State $(\hat G_1,\ldots,\hat G_{K-1})\Leftarrow (G_1 -x_1\cdot \mathrm{stepSize},\ldots,G_{K-1}-x_{K-1}\cdot \mathrm{stepSize})$
		\If{$\sho(n,\{\hat G_i^k\}_{i=1}^{K})>SE$ and $G_{\max}\geq\hat G_i\geq1\forall\ i\in\{1,\ldots,K-1\}$}
		\State $(G_1,\ldots,G_{K-1})\Leftarrow(\hat G_1,\ldots,\hat G_{K-1})$
		\Else{ $s\Leftarrow s+1$}
		\State $\mathrm{stepSize}\Leftarrow (G_{\max}-1)/2^{-s}$
		\EndIf
		\EndWhile
		\State \Return $G^K$
	\end{algorithmic}
\end{algorithm}
The complete Algorithm \ref{algo:gradient-optimizer} iterates between the two methods for a few rounds.
\begin{algorithm}[!ht]
	\caption{Gradient-based Optimization}\label{algo:gradient-optimizer}
	\begin{algorithmic}[1]
		\Require $G_{\max},\ a,\ n,\ $
		\State $(G_1,\ldots,G_{K-1})\Leftarrow(G_{\max},\ldots,G_{\max})$
		\State $SE\Leftarrow\ssh(n,G^K)$
		\While{$s<a$}
		\State $G^K\Leftarrow \mathrm{energyGradient}(G^K,s,n,SE)$
		\State $G^K\Leftarrow \mathrm{spectralSurface}(G^K,s,n,SE)$
		\EndWhile
		\State \Return $G^K$
	\end{algorithmic}
\end{algorithm}

The algorithm we developed to find approximate solutions for EGS is based on heuristic analysis and numerical study of the problem. From Methods \ref{appendix:op-gains} and Methods \ref{appendix:cont-amp} it is clear that the EGS problem is highly nontrivial in general. Our goal is thus to exploit the fact that we typically have values of $\eta$ such that $n\gg \eta^{-1}$. This results in the maximum gain obeying $G_{\max}\approx\eta^{-1}$. 
For such values of $n$ and $\eta$ we have observed that the determinant of the Hessian matrix of $\sho$, $\det(H\sho)$, is negative in a region around the gain value $(1,\ldots,1)$, while being positive for most parameter choices - in particular at $(G_{\max},\ldots,G_{\max})$. This observation justifies using $(G_{\max},\ldots,G_{\max})$ as the starting point for the algorithm. Moreover, letting the algorithm initially avoid setting a gain to one will let it effectively operate in a region where $\det(H\sho)$ is positive, so that the algorithm acts as if optimizing over a convex region. 

\begin{figure}[!t]\label{fig:contour}
	\centering
	\includegraphics[width=.45\textwidth]{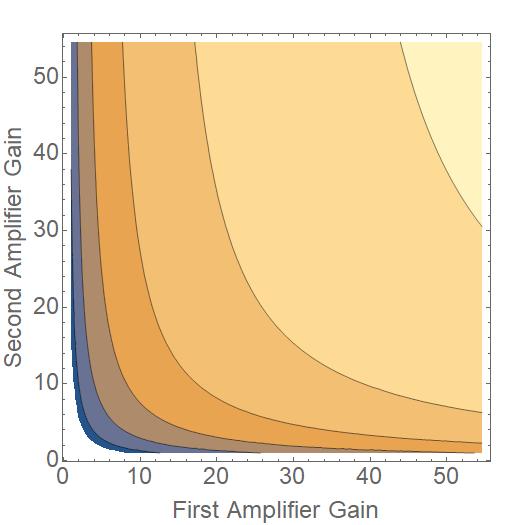}
	\includegraphics[width=.45\textwidth]{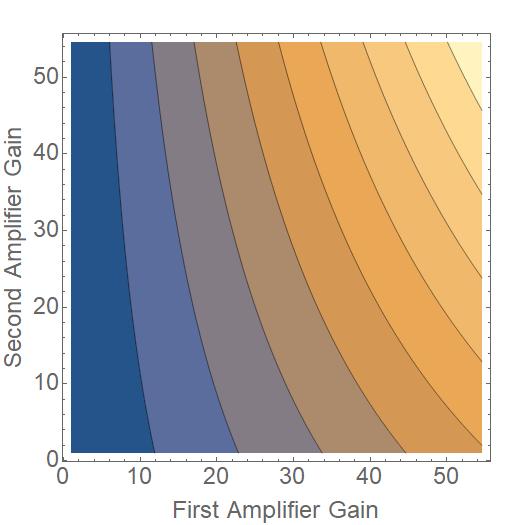}
	\caption{Displayed is a contour plot of the Holevo SE (left) and of the energy consumption (right) for $n=10^7$, $\alpha=0.05$ and $L=\SI{240}{km}$. The maximum gain in this case is $\approx54.6$. }
\end{figure}

The approach of initially moving along the gradient $\nabla E$ of $E(\cdot)$ is thus a suitable approach. The entries of $\nabla E$ are calculated based on the partial derivatives
\begin{align}
	\partial_{G_i} E(G^K)% &= \partial_{G_i}\sum_{k=1}^{K-1}(G_k - 1)(\eta(\tau_{k-1}n+\nu_{k-1})+1)\\
	%&=\partial_{G_i}\sum_{k=1}^{i-1}(G_k - 1)(\eta(\tau_{k-1}n+\nu_{k-1})+1)\\
	% &\qquad+\partial_{G_i}(G_i - 1)(\eta(\tau_{i-1}n+\nu_{i-1})+1)\\
	%&\qquad+\partial_{G_i}\sum_{k=i+1}^{K}(G_k - 1)(\eta(\tau_{k-1}n+\nu_{k-1})+1)\\
	&=\eta(\tau_{i-1}n+\nu_{i-1})+1 +\sum_{k=i+1}^{K}(G_k - 1)\eta(\partial_{G_i}\tau_{k-1}n+\partial_{G_i}\nu_{k-1}),
\end{align}
with $G^K:=\{G_i\}_{i=1}^K$.
The derivative $\partial_{G_i}\tau_{k-1}$ equals $G_1\cdot\ldots\cdot G_{i-1}\cdot G_{i+1}\cdot\ldots\cdot G_{k-1}\eta^{k-1}$. The derivatives $\partial_{G_i}\nu_{k-1}$ are calculated based on the matrix
\begin{align}
	M(G):=\left(\begin{array}{ll}\eta\cdot G& G-1\\0&1\end{array}\right)
\end{align}
which, using the canonical basis vectors $e_1,e_2$, lets us compute each $\nu_k$ as
\begin{align}
	\nu_k &= \langle M(G_k)\cdot\ldots\cdot M(G_1)e_2,e_1\rangle.
\end{align}
Denoting with $D(G):=\partial_G M(G)$ the derivative of $M$ at point $G$, the derivative of $\nu_k$ with respect to $G_i$ is (if $i<k$)
\begin{align}
	\partial_{G_i}\nu_k(G^k) &= \langle M(G_k)\ldots M(G_{i+1})\cdot D(G_i)\cdot M(G_{i-1}) \ldots M(G_1)e_2,e_1\rangle.
\end{align}
For the large values of $n$ studied in this work the approximation $G_{\max}\approx\eta^{-1}$ holds. At the point $G^K=(\eta^{-1},\ldots,\eta^{-1},1)$ however we have (for $i<k$) the special situation that \begin{align}
	M(\eta^{-1})^{i-1}&=\left(\begin{array}{ll}1&(i-1)(\eta^{-1}-1)\\0&1\end{array}\right)\\
	D(\eta^{-1}) M(\eta^{-1})^{i-1}&=\left(\begin{array}{ll}\eta&(i-1)(1-\eta)+1\\0&0\end{array}\right).
\end{align}
Further it holds $M(\eta^{-1})^{k-i+1}D(\eta^{-1}) M(\eta^{-1})^{i-1}=D(\eta^{-1}) M(\eta^{-1})^{i-1}$, so that
\begin{align}
	\partial_{G_i}\nu_k(G^k) &= i - (i-1)\eta,\\
	\partial_{G_i}\tau_{k-1}n &= \eta n,\\
	\nu_{i-1} &= (i-1)(\eta^{-1}-1).
\end{align}
Thus since $\eta<1$ we have $x:=\tfrac{1}{\eta}-1>0$  and can lower bound the components of the gradient $\nabla E$ as
\begin{align}
	\partial_{G_i} E(G^K) &= \eta(n+(i-1)x)+1+\sum_{k=i+1}^Kx\cdot \eta^2(n+1 + i\cdot x)\\
	&\geq\eta\cdot n+ 1.
\end{align}
Since $K$ is at most $20$ in all the situations we investigated, $n\gg x$ and $\eta\gg\eta^2$, the above lower bound is approximately tight within the limits of our heuristic analysis. A typical situation for two gains is presented in Figure \ref{fig:contour}.
From above estimates we conclude the initial step of our optimization will typically avoid reducing one of the gains drastically, so that the algorithm effectively behaves as if optimizing over a convex region where $\det(H\sho)>0$.\\
Based on the available equations \eqref{eqn:der-hol} for the derivatives of the Holevo SE, $\nabla \sho$ is available as well.  
%The Python code for the optimization procedure is attached to this manuscript.

\subsection{Shannon performance with homodyne detection}\label{appendix:homodyne}
In~\cite{Takeoka14} it was shown that the best Gaussian receivers for communication on lossy bosonic channels (including the one studied in this paper) are heterodyne, homodyne or a time-sharing between the two. The transition between the heterodyne-optimal and homodyne-optimal regimes takes place as the energy of the attenuated signal decreases, around $\tau\cdot n\approx 2$. Moreover, since the OSSR for Gaussian channels has been recently proved to be Gaussian~\cite{Holevo2019}, we conclude that the hetero/homodyne receiver coincides with the OSSR for our channel. 

\begin{figure}[!t]\label{fig:homodyne_egs}
	\centering
	\includegraphics[width=.95\textwidth,valign=t]{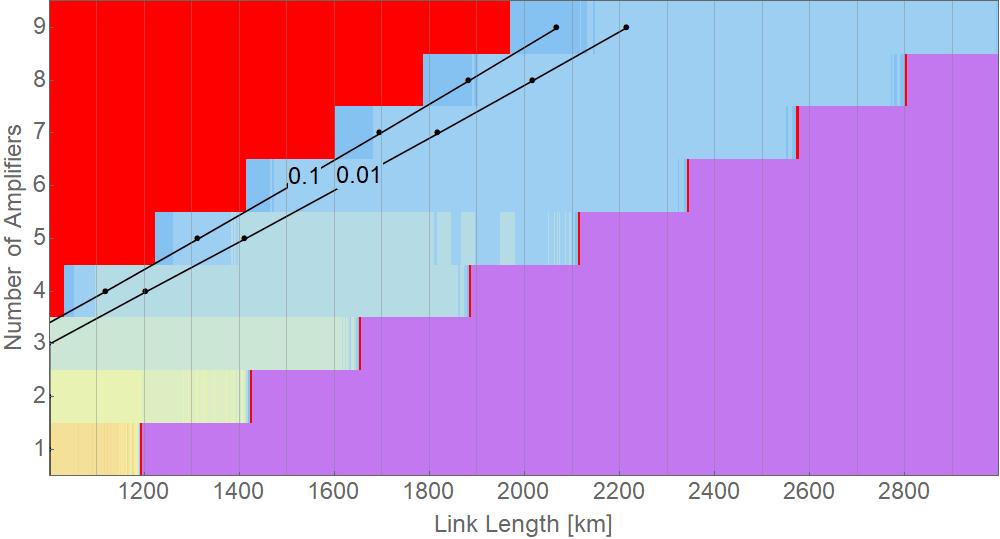}
	\includegraphics[width=\textwidth,valign=t]{assets/legendRow.jpg}
	\caption{Displayed are the energy savings achievable with homodyne OSSR over heterodyne OSSR. In the red region, the homodyne receiver is inferior to the heterodyne one. In the violet region, the savings are above $95\%$.}
\end{figure}
In the main text we focused on comparing the performance of the heterodyne receiver with the OJDR, observing a striking energy advantage of the latter. 
Here instead we briefly compare the performance of the homodyne receiver with the OJDR. The Shannon SE attained by using the homodyne receiver is~\cite{Takeoka14}
\begin{equation}
	S_{\rm sh}^{\rm hom}(n):=\log\left(1+\frac{4\tau\cdot n}{1+2\nu}\right).
\end{equation}
For this quantity, the SEGS gain profile can in general be different from full amplification, similarly to the Holevo SE. Indeed, defining the signal-to-noise ratio as $\snr':=4\tau\cdot n/(1+2\nu)$ and using the same notation as Methods~\ref{appendix:op-gains}, we have
\begin{align}
	\frac{\partial \snr'}{\partial G_i}&=\frac{\snr'}{G_i}-\frac{\snr'}{1+2\nu}\cdot2\cdot \frac{\nu-\nu_{>i}+\tau_{>i}}{G_i}\nonumber\\
	&=\frac{\snr'}{G_i}\cdot\frac{1-2\beta_{i}}{1+\nu}.\label{eq:derSnr2}
\end{align}
Since $\beta_i\leq1$ for all $i$, this derivative can be negative and hence we can conclude that in general the homodyne SE will be maximized my non-maximum gains. In particular, if this holds in a region where the homodyne receiver can surpass the heterodyne receiver, we can expect the former to also offer an energy advantage with respect to the heterodyne receiver.
For this reason, we also analyzed the EGS problem when using a homodyne receiver instead of the OJDR. The results, shown in Figure~\ref{fig:homodyne_egs}, confirm that the advantage of OJDR is much more pronounced, particularly in the practically relevant region identified in the main text. For example, in the same setting where the OJDR has record savings of $55\%$, the homodyne detector offers no savings at all - its SE is below $\SI{8}{bits/s/Hz}$, while that of the fully amplified link with OSSR reach a value of above $\SI{14}{bits/s/Hz}$.

%\section{Conclusions}

\backmatter

%\bmhead{Supplementary information}

\bmhead{Acknowledgments} Matteo Rosati gratefully acknowledges Marco Fanizza, for useful discussions of the results and for pointing out the use of data-processing to remove the last amplifier, as well as Andreas Winter, for useful discussions on the gap between Holevo and Shannon capacities.
Janis Nötzel gratefully acknowledges stimulating discussions on the topic with Marcin Jarzyna and Konrad Banaszek during his visit at the CeNT.\\
This project has received funding from the DFG Emmy-Noether program under grant number NO 1129/2-1 (JN) and the European Union’s Horizon 2020 research and innovation programme under the Marie Skłodowska-Curie grant agreement No 845255 (MR). Janis Nötzel further acknowledges the financial support by the Federal Ministry of Education and Research of Germany in the programme of "Souverän. Digital. Vernetzt.". Joint project 6G-life, project identification number: 16KISK002, and of the Munich Center for Quantum Science and Technology (MCQST). 
Both authors thank the participants of EACN 2022, and in particular Dan Kilper, for pointing out the interesting properties of hollow-core fiber.
\bmhead{Declarations}
The authors declare no competing interests.

%\begin{appendices}
%\end{appendices}

\bibliography{bib}

\end{document}